\documentclass[aps,prl,preprint,groupedaddress]{revtex4-1}

\usepackage[margin=1.0in]{geometry}
\usepackage{verbatim}
\usepackage{graphicx}
\usepackage{amsmath}
\usepackage{epstopdf}
\usepackage{color}
\usepackage{setspace}

\begin{document}


\title{Observation of Multimode Solitons in Few-Mode Fiber}

\author{Zimu Zhu,$^{1,}$\footnote{zz346@cornell.edu}
Logan G. Wright,$^{1}$
Demetrios N. Christodoulides,$^{2}$
Frank W. Wise$^{1}$
} 
    \affiliation{
	$^1$School of Applied and Engineering Physics, Cornell University, Ithaca, New York 148853, USA
	}
	\affiliation{
	$^2$CREOL, The College of Optics and Photonics, University of Central Florida, Orlando, Florida 32816, USA
	}

\begin{abstract}
We experimentally isolate and directly observe multimode solitons in few-mode graded-index fiber. By varying the input energy and modal composition of the launched pulse, we observe a continuous variation of multimode solitons with different spatiotemporal properties. They exhibit an energy-volume relation that is distinct from those of single-mode and fully spatiotemporal solitons.
\end{abstract}


\maketitle

In single mode fiber (SMF), soliton solutions of the (1+1)D nonlinear Schr\"{o}dinger equation (NLSE) occur when linear dispersion and the Kerr nonlinearity balance to produce localized pulses \cite{Hasegawa1973, Mollenauer1980}. While the literature on (1+1)D soliton dynamics is extensive \cite{AgrawalNLFO}, nonlinear dynamics in multimode fiber (MMF) is less explored. In MMF, multiple spatial eigenmodes are supported, with intra- and inter-modal contributions to the dispersion, and propagation involves both spatial and temporal degrees of freedom \cite{Crosignani1982, Mafi2012}. Although theoretical predictions of multimode solitons date back to the 1980s \cite{Hasegawa1980, Crosignani1981, Yu1995, Chien1996, Raghavan2000, Lederer2008}, few experiments on multimode solitons have been reported. Renninger {\textit{et al.}} first reported solitons in graded-index (GRIN) MMF, but the solitons were dominated by the fundamental mode \cite{Renninger2013}. Wright {\textit{et al.}} followed up on this work by exploring different regimes of spatiotemporal dynamics in MMF \cite{Wright2015}. Wright {\textit{et al.}} presented evidence of solitons consisting of up to about 10 modes, but due to the many fiber modes, interpretation of the experimental measurements was complicated. Thus, experiments that add systematic understanding of MM solitons are needed. In particular, the study of solitons in a small number of modes is a natural step in the progression from single-mode to many-mode, and ultimately, bulk spatiotemporal systems. Few-mode fiber (FMF) is the perfect platform for such studies. Buch {\textit{et al.}} recently numerically observed trapping of SM temporal solitons in the different modes of a FMF \cite{Buch2015}. Here, we aim to experimentally observe MM solitons in FMF.

Apart from their intrinsic interest, MM solitons are relevant to a variety of applications. There is interest in MMF for higher-capacity data transmission through space-division multiplexing (SDM) \cite{Ryf2012, Richardson2013, Bordague1982, Stuart2000, Tarighat2007}. As an example, transmission over 1,000 km by SDM in three-mode fiber was demonstrated with appropriate amplification \cite{Esmaeelpour2016}. The larger mode areas of MMFs serve as a route to fiber lasers with higher power or pulse energy \cite{Fermann1998}. More broadly, knowledge of the properties of MM solitons may provide a basis for understanding complex nonlinear dynamics in MMF, in the same way that knowledge of (1+1)D solitons does for SMF \cite{WrightPRL}.

\begin{figure}
\fbox{\includegraphics[width=10cm]{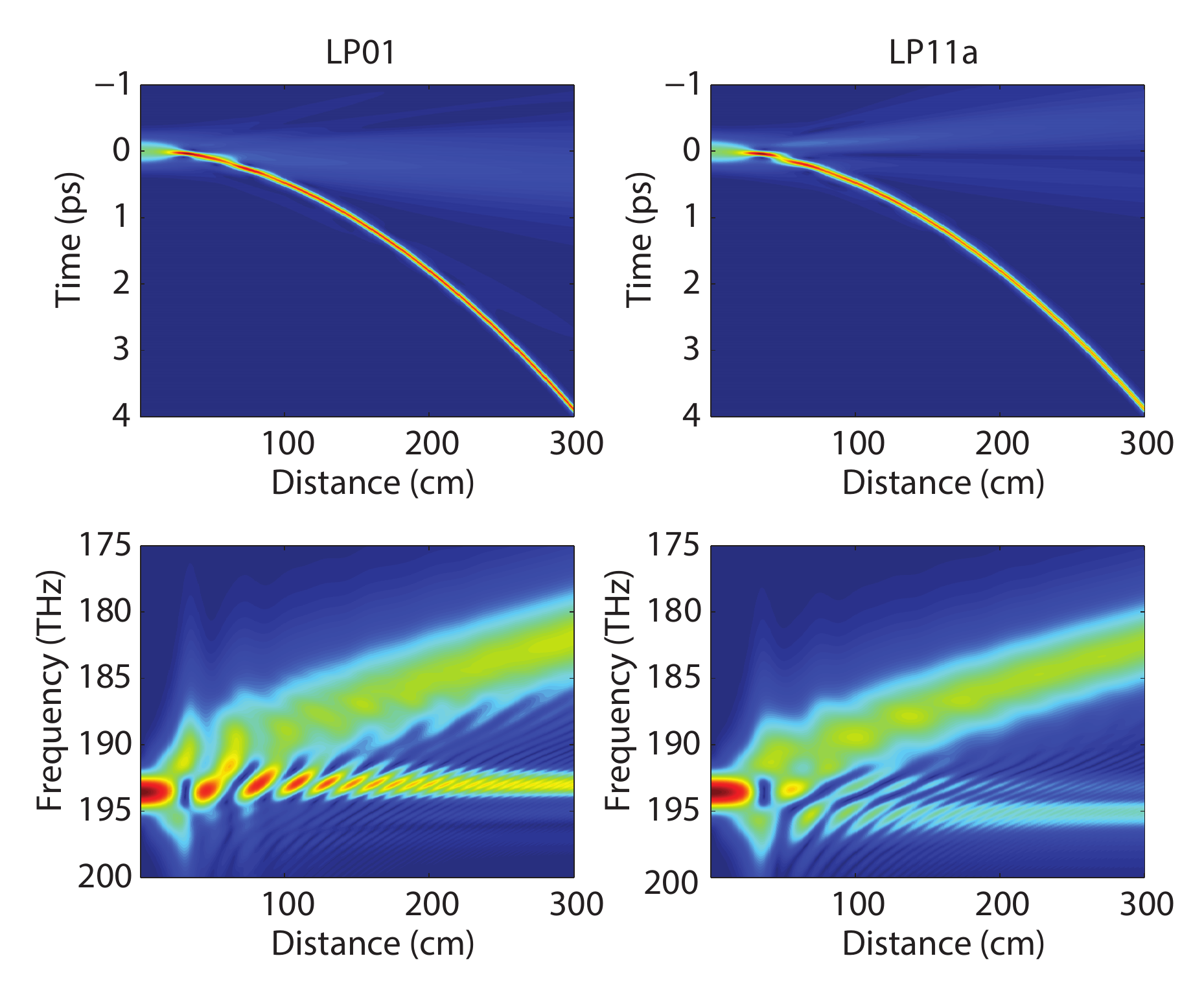}}
\caption{Simulated temporal and spectral evolutions of the LP01 and LP11a modes over a propagation distance of 3 meters, for 5 nJ total energy initially spread equally over the three modes. The LP11b mode follows a nearly identical evolution to that of LP11a.}
\label{fig_sim_evol}
\end{figure}

In this letter, we report the results of pulse-propagation experiments designed to isolate MM solitons in graded-index (GRIN) fiber that supports only the LP01 and LP11 mode groups (three spatial eigenmodes in total). We focus on solitons that have red-shifted as a result of Raman scattering, which allows them to be spectrally isolated in an experimentally straightforward way that is lacking in prior experimental studies. We observe that variation in the initial launch conditions (pulse energy and mode content) leads to solitons with a continuous range of spatiotemporal properties. We determine the relationship between soliton energy and duration, as well as the variation in modal content with pulse energy. Increased energy of MM solitons is accommodated by adjustments in both the spatial and temporal domains. The experimental results agree reasonably with numerical simulations. 

The fiber used in our experiment is a commercial product (Two Mode Graded-Index Fiber from OFS, Inc.), with a proprietary design. However, we can estimate its parameters from the available information. We assume a GRIN fiber with a parabolic index profile, a numerical aperture of 0.14 and a core diameter of 16 um. Such a fiber supports the LP01 and LP11 mode groups with low differential group delay ($<$ 0.2 ps/m). 

To guide the experiments, and to derive detailed insight from the results, we first simulated the pulse propagation using the generalized multimode nonlinear Schr\"{o}dinger equation (GMM-NLSE) \cite{Wright2015, Poletti2008}. This is the same method used by Wright \textit{et. al.} in Ref. \cite{Wright2015}. For fibers with a small number of modes, this method provides detailed information about the individual modal evolutions, while remaining computationally inexpensive. In our simulations, we launched a 250-fs Gaussian pulse centered at 1550 nm, and monitored the propagation over a distance of 6 m for a variety of different initial conditions. Total pulse energy was varied, along with the modal composition, by varying the absolute and relative peak intensities of the modes in the GMM-NLSE.  


\begin{figure}
\fbox{\includegraphics[width=10cm]{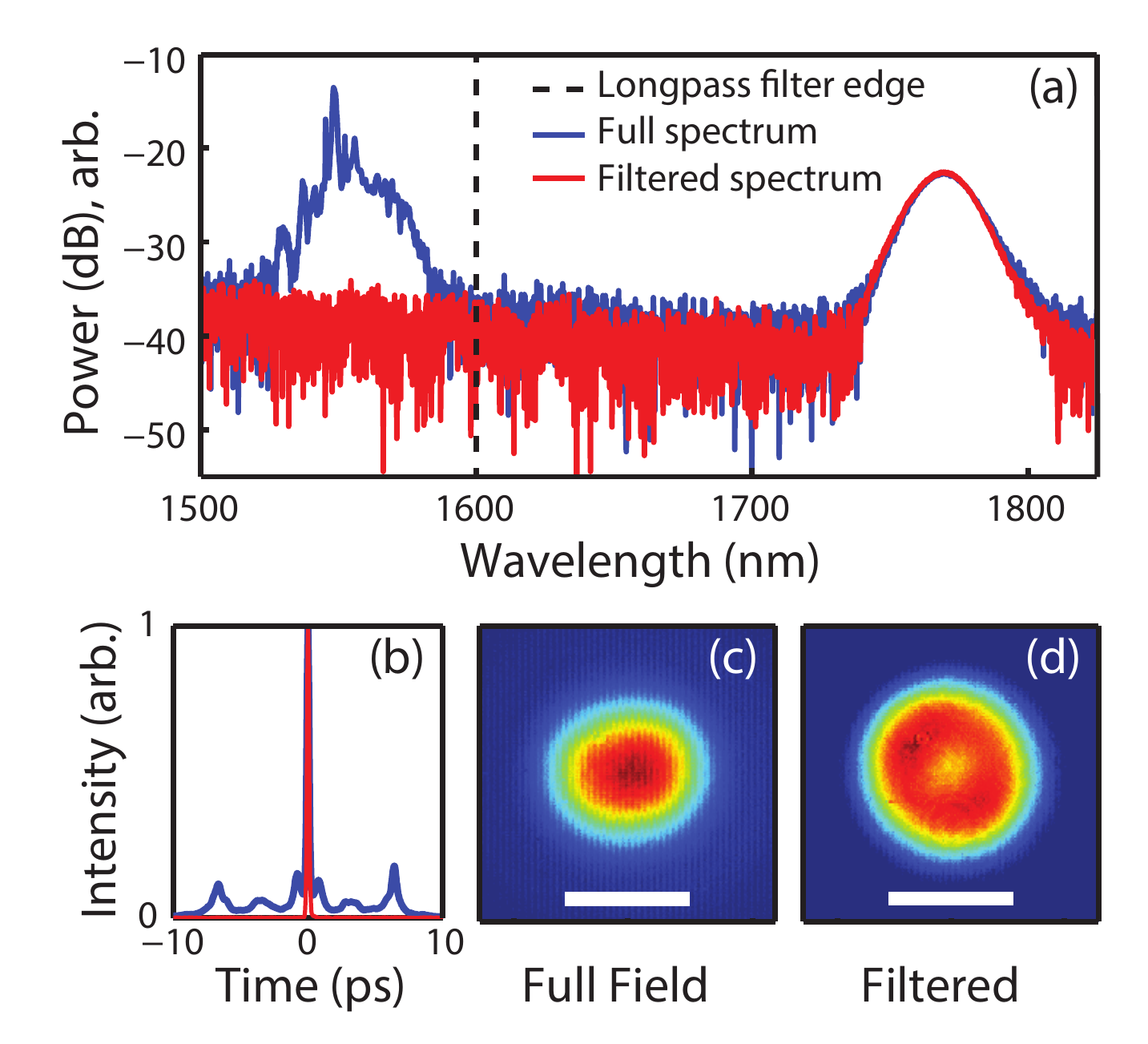}}
\caption{Sample data (blue lines: full spectrum, red lines: spectrum transmitted by 1600-nm long-pass filter) for 9-nJ pulses launched into 25 m of FMF: (a) spatially-integrated spectra (dashed line represents long-pass filter edge), (b) intensity autocorrelation traces, (c) spectrally-integrated beam profile for full spectrum, (d) spectrally-integrated beam profile for $\lambda > 1600$ nm (white bar represents 11 $\mu$m mode-field diameter of the fundamental mode of the fiber).}
\label{fig_exp_example}
\end{figure}

The simulation results provide a clear picture of how MM solitons form. In the linear propagation regime, individual modes broaden and separate temporally due to group velocity dispersion (GVD) and intermodal dispersion respectively. At higher energies, where the nonlinear length becomes comparable to the dispersion length and shorter than the walk-off length (the length scale on which the modes lose their temporal overlap due to having differing modal group velocities), MM solitons form. A particular example is shown in Fig.~\ref{fig_sim_evol}. For a total energy of 5 nJ spread equally across the modes, compression of the initial pulse leads to MM soliton formation through a fission process. The pulse sheds energy, emitting dispersive radiation in the modes (not locked together in time) that can be observed in the figure. The pulse then undergoes temporal periodic breathing in its width and peak intensity as the three modes exchange energy with one another, which diminishes as it approaches a stable soliton product. The pulse spectrum (bottom panels of Fig.~\ref{fig_sim_evol}) shifts to longer wavelengths owing to stimulated Raman scattering \cite{Wright2015}, which provides the necessary shifts in the modal group velocities for the individual modes to lock together at a common group velocity. The result is a multimode (3-mode) Raman soliton, with spectral characteristics similar to those of Raman solitons in SMF \cite{Mitschke1986}. While intermodal Raman scattering is not necessary for the existence of MM solitons, as intermodal cross-phase modulation (XPM) and four-wave mixing (FWM) can also provide the necessary nonlinear phase and group velocity shifts, it plays a key role in assisting their formation. As a result, spectral shifts due to Raman scattering are a dominant effect in MM solitons even with moderate pulse energies. 

Experimentally, we launched 280-fs pulses at 1550 nm, with energies ranging up to 20 nJ, into 25 m of the FMF. The fiber corresponds to 8 dispersion lengths for the input pulse. The output from the fiber was measured in the spatial, spectral, and temporal domains using an InGaAs camera, spectrum analyzer and intensity autocorrelator, respectively. While there currently exist spatiotemporal diagnostics such as the STRIPED-FISH technique \cite{Trebino2016}, it has not yet been realized with sufficient time-bandwidth product to be suitable for measuring the pulses in our experiment. A half-wave plate and polarizing beamsplitter were used to vary the input pulse energy. The launch condition was controlled by fixing the bare input end of the FMF to a three-axis translation stage, which could be moved relative to the input beam focus. Although this does not provide the capability to exactly control or measure the modal composition of the launched pulses, we are still able to sweep through a large parameter space and to intentionally excite higher-order mode content. Spherical and cylindrical lenses were used to help deliberately excite the LP11 group. 

While soliton formation is most naturally studied in a cutback experiment, such a scheme is not feasible with MMF, since any mechanical manipulation of the fiber may modify the input alighment and possibly lead to changes in the linear mode coupling. Thus, we kept the fiber length fixed, and measured the output for varied initial conditions. To observe MM Raman solitons, we looked for three experimental signatures: the shifted spectrum, unbroadened temporal profile, and multimode beam profile. A 1600-nm longpass filter allows us to isolate the Raman-shifted soliton from the rest of the field, so it can be directly measured in all three domains. In this work we intentionally ignore low energy regimes in which MM solitons form without significant Raman scattering, since we are unable to separate a single MM-soliton from the rest of the field. 

\begin{figure}
\fbox{\includegraphics[width=10cm]{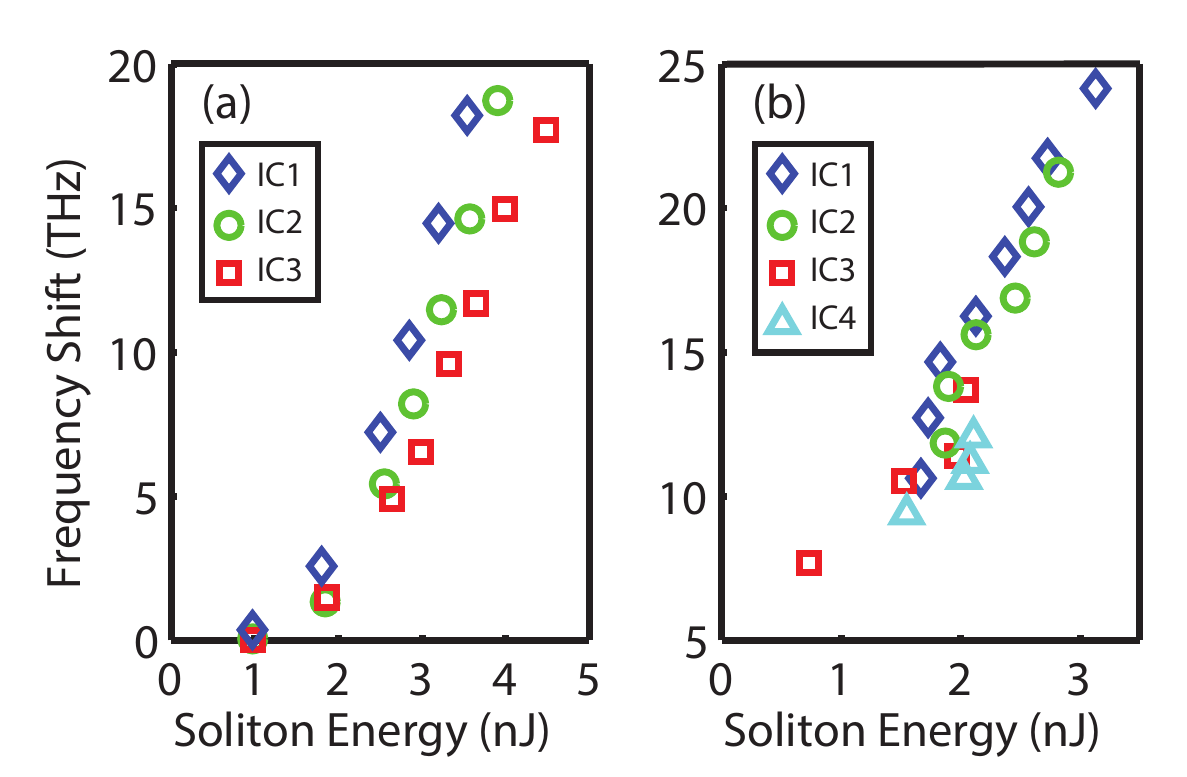}}
\caption{Raman frequency shift of MM solitons versus soliton pulse energy in: (a) simulation and (b) experiment ($\lambda > 1600$ nm). Colours and symbols denote different initial conditions (IC). IC1: LP01 dominated, IC2: intermediate, IC3: LP11 dominated, IC4: LP11 dominated.}
\label{fig_sim_trends}
\end{figure}

A representative set of data that demonstrates MM soliton formation is presented in Fig.~\ref{fig_exp_example}. These particular data are the result of launching 9-nJ pulses into the FMF with an initial fundamental mode-dominated excitation (a reasonable estimate of the modal excitation can be obtained by looking at the fiber output at low energy). The spectrum exhibits a clear Raman-shifted peak, along with dispersive radiation left behind around 1550 nm. Using the filter, a 120-fs Raman soliton can be spectrally isolated from the full field (Fig.~\ref{fig_exp_example}(a)) and observed in the autocorrelation trace (Fig.~\ref{fig_exp_example}(b)), in which both traces are normalized to unity intensity at $t = 0$. Given that for low energy pulses (1 nJ), the measured output pulse width is roughly 1.6 ps, these short Raman pulses are the product of nonlinear processes that have counteracted pulse broadening mechanisms. This spectral filtering is done in the spatial domain as well, allowing us to observe the spatial variation between the full field (Fig.~\ref{fig_exp_example}(c)) and the Raman soliton above 1600 nm (Fig.~\ref{fig_exp_example}(d)). Thus we are able to isolate and observe individual MM solitons in the spectral, temporal and spatial domains.

\begin{figure}
\fbox{\includegraphics[width=10cm]{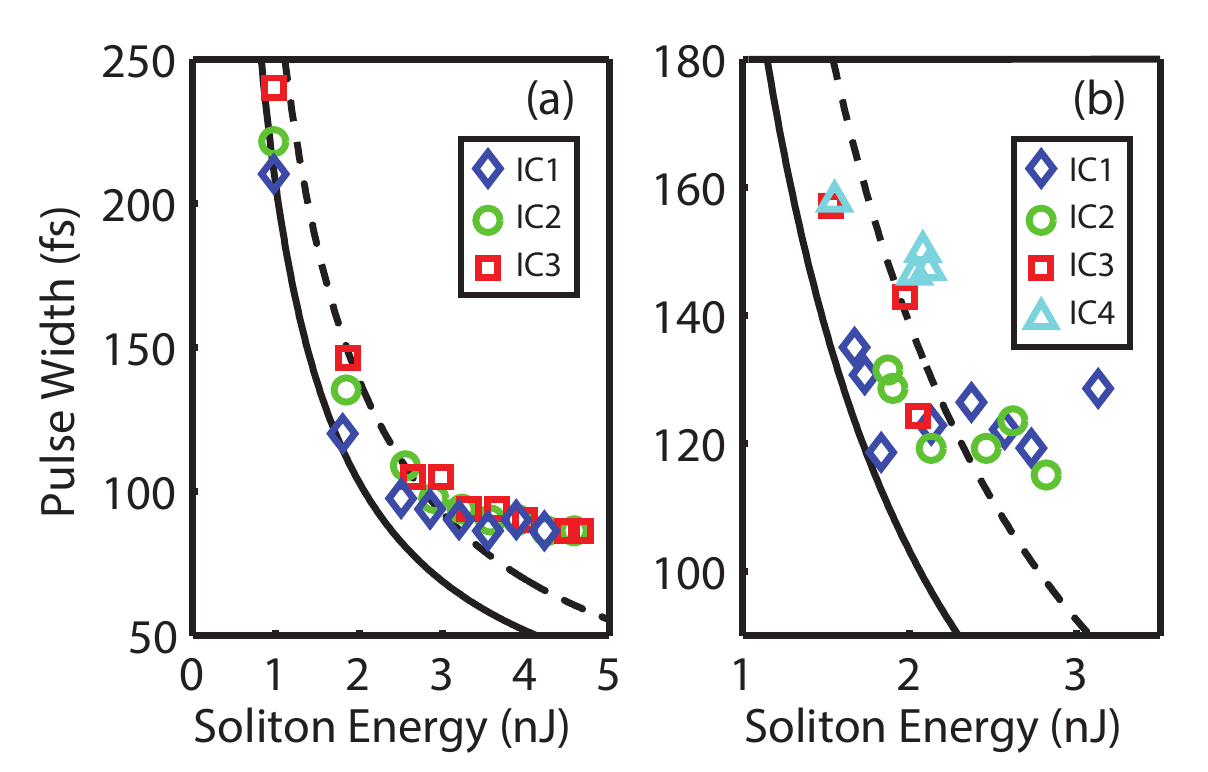}}
\caption{Pulse width of MM Raman soliton versus energy in: (a) simulation and (b) experiment ($\lambda > 1600$ nm). Colours and symbols denote different initial conditions (IC). IC1: LP01 dominated, IC2: intermediate, IC3: LP11 dominated, IC4: LP11 dominated. Solid line depicts soliton energy area theorem for LP01 mode. Dashed line depicts soliton energy area theorem for LP11 mode.}
\label{fig_exp_trends}
\end{figure}

We observe that, for energy ranges in which a single MM Raman soliton forms, the resulting modal composition of the MM soliton is similar to that of the launched pulse. As a result, the Raman frequency shift that each MM soliton experiences lies between the frequency shifts that would be experienced by a similar pulse with the same total energy in either the fundamental or higher order mode. This is displayed in Fig.~\ref{fig_sim_trends}, which shows the same trend for both the numerical and experimental data. A single-mode soliton in the fundamental mode would experience the greatest Raman shift, and a lesser shift would be experienced by a single-mode soliton in an LP11 mode. In Fig.~\ref{fig_sim_trends} and Fig.~\ref{fig_exp_trends}, the blue diamonds correspond to initial pulses dominated by the fundamental mode, and as such they experience the greatest frequency shift. Conversely, the red squares, as well as the cyan triangles in Fig.~\ref{fig_sim_trends}(b) and Fig.~\ref{fig_exp_trends}(b), are different initial conditions both dominated by the LP11 mode group, and as such experience a smaller shift. Similar trends hold for the pulses’ temporal delay and group velocity, which is discussed in the next paragraph. The importance of this observation is that it demonstrates the variability in the properties of MM solitons as a multi-parameter family of solutions to the GMM-NLSE, a characteristic that is different from SM solitons. In our simulations, we find that at even higher initial pulse energies, these trends no longer hold true: stronger soliton fission leads to multiple MM soliton products with less predictable characteristics. More work is required to systematically investigate this regime.

\begin{figure}
\centering
\fbox{\includegraphics[width=10cm]{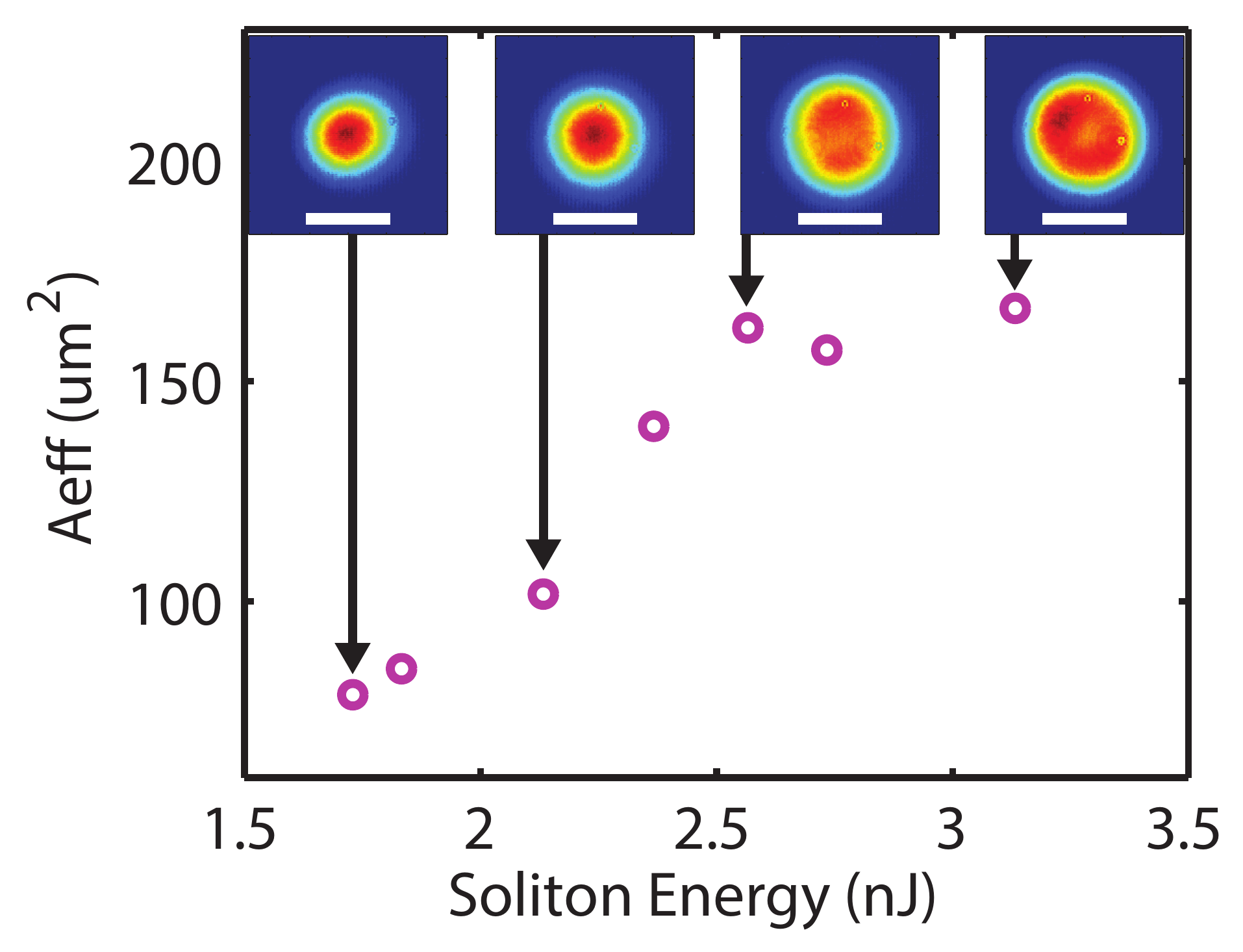}}
\caption{Effective mode area as a function of energy for a given initial launch condition, calculated as $A_{\text{eff}} = (\int{I~ \mathrm{d} A})^2/\int{I^2~ \mathrm{d} A}$. Inset: beam profiles corresponding to the data points (white bar represents an 11 $\mu$m scale for reference).}
\label{fig_trend_beamsize}
\end{figure}

We find that MM solitons are able to adjust spatiotemporally to an increase in energy via transverse/modal degrees of freedom in a way that is impossible for (1+1)D solitons in SMF. With multiple modes supported, one would suspect that MM solitons are capable of having higher pulse energies than SM solitons. This is confirmed by both our experimental and numerical results in Fig.~\ref{fig_exp_trends}, which shows pulse duration as a function of pulse energy. At low energies, before any significant Raman frequency shift occurs, the pulse behaves like a superposition of the modes. As a result, at these energies (up to 2.5 nJ), the pulse width lies between the soliton energy area relations for the LP01 and LP11 modes \cite{AgrawalNLFO}. This is most evident in Fig.~\ref{fig_exp_trends}(a), where again the fundamental-mode-dominanted pulses (blue diamonds) lie closer to the solid black line (LP01), while the higher-order-mode-dominated pulses (red squares) lie closer to the dashed line (LP11). The same is true in Fig.~\ref{fig_exp_trends}(b). As energy increases, the MM soliton duration decreases, converging toward a fairly constant duration of about 80 fs in simulation and 120 fs in experiment. Here the MM soliton energies exceed what's predicted by the single-mode soliton energy area relations for both the LP01 and LP11 modes. In other words, transverse spatial degrees of freedom allow the MM soliton to contain more energy than any SM soliton could in any of the individual modes of the fiber. Moreover, we find that as energy increases, the spatial size of the pulse increases as well (while maintaining a roughly constant pulse duration). Fig.~\ref{fig_trend_beamsize} shows this for one initial condition, with a similar trend observable across all data recorded. The initial beam profile of the MM soliton closely resembles the fundamental mode, and grows in size as energy is increased. The effective area then plateaus as the beam fills the fiber. From there, subsequent adjustments to further energy increase are made through changes in modal content.

It is worth noting that not only do MM solitons behave differently from SM solitons, but they behave differently from fully spatiotemporal solitons as well (in bulk material or free space). While (3+1)D spatiotemporal solitons are expected to decrease in both temporal and spatial size as energy is increased \cite{Silberberg1990}, MM solitons on the other hand are like temporal solitons with spatial degrees of freedom - spatiotemporal objects that decrease in temporal size but increase in spatial size. As such, it is clear that MM solitons in systems with discrete modes occupy a qualitatively different space between that of (1+1)D single-mode solitons and (3+1)D spatiotemporal solitons that are not characterized by a finite number of modes. In this way, further investigation of MM solitons could shed light on the transition of the spatiotemporal properties of solitons from one-dimensional systems to fully three-dimensional ones.

In summary, we have experimentally isolated and observed multimode solitons in GRIN FMF, and find them to display a continuous range of spatiotemporal properties that are dependent on their modal composition. These MM solitons are stable at higher energies than is otherwise possible in SMF, a property attributed to the transverse spatial degrees of freedom that result from the fiber supporting multiple spatial eigenmodes. They display an energy-volume relation that is different from those of SM solitons and fully spatiotemporal solitons. In the future it will be desirable to perform direct mode-resolved measurements of MM solitons.  Analogous studies of MM solitons in step-index and multicore fibers will also be interesting.\\ 

Portions of this work were funded by Office of Naval Research grant N00014-13-1-0649. L. G. W. and Z. Z. acknowledge support from NSERC. We thank OFS for providing the fibre used in the experiments at a discount. We also thank Corning for other fibers used in the experiments.

\end{document}